\documentclass[twocolumn,showpacs,amsmath,amssymb,longbibliography]{revtex4-1}
\usepackage{bm}
\usepackage[T1]{fontenc}
\usepackage{hyperref}
\input{epsf}

\usepackage{graphicx}
\usepackage{epstopdf}
\usepackage{array}
\usepackage{longtable}
\usepackage{rotating,booktabs}
\usepackage{booktabs,threeparttable}
\usepackage{bm}
\usepackage{float}
\usepackage{amsmath}
\usepackage{gensymb}
\usepackage{multirow}
\usepackage{epsfig}
\usepackage{color}

\begin{document}

\title{\emph{Ab initio} calculation of hyperfine-structure properties to extract nuclear magnetic octupole moments of $^{69}$Ga and $^{71}$Ga}
\author{Fei-Chen Li$^1$ and Yong-Bo Tang$^{2,*}$}

\affiliation{$^1$School of Physics and advanced energy, Henan University of Technology, Zhengzhou 450001, China}
\affiliation {$^2$College of Engineering Physics, Shenzhen Technology University, Shenzhen, 518118, China}
\email{tangyongbo@sztu.edu.cn}
\date{\today}

\begin{abstract}
We calculate the hyperfine-structure properties of the low-lying states in $^{69}$Ga and $^{71}$Ga using the relativistic coupled-cluster method at the singles and doubles (RCCSD) level. The properties include the first-order hyperfine constants and the second-order corrections arising from off-diagonal hyperfine interactions.
Based on our theoretical results, we reanalyze the previous hyperfine-splitting measurements of the $4p_{3/2}$ state in $^{69,71}$Ga [R. T. Daly and J. H. Holloway, Phys. Rev. 96, 539 (1954)], and refine the nuclear magnetic octupole moments of these two isotopes. The refined values $\Omega(^{69}\mathrm{Ga})=0.123(9)\,\mu_N\times b $ and $\Omega(^{71}\mathrm{Ga})=0.165(15)\,\mu_N\times b $ are about 13\% and 12\% larger than the earlier results reported by Daly and Holloway. By providing the first independent \emph{ab initio} determination, the present results offer a reliable set of reference values for the magnetic octupole moments of $^{69,71}$Ga, which will benefit both future high-precision measurements and nuclear-structure calculations.
\end{abstract}
\maketitle

\section{Introduction}
\label{sec:introduction}

The atomic hyperfine structure (HFS) arises from the interaction between nuclear electromagnetic multipole moments and the electron cloud~\cite{ Schwartz1955}, serving as an important bridge between atomic and nuclear physics and providing a valuable means for determining nuclear moments~\cite{Bieron2005, Bieron2009}. While the magnetic dipole (M1) and electric quadrupole (E2) moments have been systematically established by combining various experimental methods with atomic and molecular theoretical calculations for many nuclei~\cite{Pyykko2018}, the magnetic octupole (M3) moment, being typically several orders of magnitude smaller, has long been neglected. Only recently, with advances in laser spectroscopy and precise atomic structure calculations, have magnetic octupole interactions begun to receive experimental and theoretical scrutiny in a limited number of atomic systems~\cite{Gerginov2003, Xiao2020, deGroote2021,Beloy2008}, but systematic studies remain scarce for most nuclei. The difficulty lies in the demanding atomic-structure description required to extract $\Omega$ from hyperfine intervals. Yet $\Omega$ contains valuable high-order information on nuclear current and magnetization distributions, and its precise determination is therefore essential for constraining the poorly known isoscalar component of nuclear spin-spin forces~\cite{Beloy2008} and for reducing uncertainties in nuclear Schiff moments~\cite{Dobaczewski2018} , placing it at the forefront of current research at the atomic-nuclear interface.

Gallium (Ga, $Z=31$) is an ideal system for investigating magnetic octupole interactions. Both stable isotopes, $^{69}$Ga and $^{71}$Ga, have nuclear spin $I=3/2$ and sizable magnetic dipole moments~\cite{Stone2005}. Their ground configuration $4s^2 4p$ yields the ground state $4p_{1/2}$ and the metastable $4p_{3/2}$, the latter being the key state for studying magnetic octupole interactions. As early as 1954, Daly and Holloway employed atomic-beam magnetic resonance techniques to precisely measure the hyperfine splitting of the $4p_{3/2}$ state in $^{69,71}$Ga. With the aid of theoretical corrections provided by C.~Schwartz, they extracted the hyperfine constants $A$, $B$, and $C$ from the experimental data, and, for the first time, evaluated the nuclear magnetic octupole moments~\cite{Daly1954}. While $A$ and $B$ were essentially determined by experiment, the extraction accuracy of the tiny $C$ constant (on the order of $0.1$\,kHz) was appreciably limited by the semiempirical theoretical corrections available at the time. Likewise, owing to the theoretical limitations of that era, the magnetic octupole moments they assessed carry substantial uncertainties.

Over the past seven decades, considerable progress has been made in both experimental and theoretical studies of the gallium HFS. On the experimental side, Neijzen and D\"onszelmann~\cite{Neijzen1980} and J\"onsson \emph{et al.}~\cite{Jonsson1985} measured the hyperfine structure and isotope shifts of Ga using laser spectroscopy techniques; modern collinear laser spectroscopy has further extended measurements to radioactive $^{67-82}$Ga isotopes, systematically determining the evolution of nuclear spins, magnetic dipole moments, and electric quadrupole moments, and revealing abrupt structural changes between $N=40$ and $N=50$~\cite{Cheal2010, Mane2011, Procter2012, deGroote2017a,deGroote2017b}. On the theoretical side, earlier work by Koster~\cite{Koster1952} explored the effects of configuration interaction on the hyperfine structure of Ga, while Sternheimer~\cite{Sternheimer1972} calculated the quadrupole shielding correction for Ga. These pioneering studies laid the foundation for later high-precision calculations. Tokman \emph{et al.} performed large-scale multiconfiguration Hartree--Fock (MCHF) calculations and accurately determined the nuclear quadrupole moments of $^{69,71}$Ga~\cite{Tokman1998}; Safronova \emph{et al.} computed the hyperfine constants $A$ for numerous excited states of Ga~I using the relativistic all-order method, which is equivalent to the linearized coupled-cluster singles and doubles approach~\cite{Safronova2006}; Wang and Dong systematically investigated the hyperfine constants $A$ and $B$ of $^{71}$Ga employing the multiconfiguration Dirac--Fock (MCDF) method~\cite{Wang2012a, Wang2012b}. However, all these theoretical studies are restricted to the magnetic dipole and electric quadrupole terms; the magnetic octupole term has never been included in any high-precision \emph{ab initio} calculation.

Recently, we carried out similar relativistic coupled-cluster calculations for indium (In, $Z=49$), which belongs to the same group IIIA, and found that second-order hyperfine effects significantly influence the extraction of the In magnetic octupole moment~\cite{Li2024}. Compared with indium, gallium has a smaller nuclear spin ($I=3/2$), fewer hyperfine levels, and only three measurable intervals, making the extraction of the $C$ constant more heavily dependent on theoretical corrections. More importantly, the determination of the magnetic octupole moment also relies on the accuracy of the diagonal matrix element $C/\Omega$, which is highly sensitive to electron correlation effects and requires high-precision \emph{ab initio} calculations for a reliable evaluation. Therefore, a systematic \emph{ab initio} study of the magnetic octupole moment of gallium is particularly necessary.

In this work, we employ the relativistic coupled-cluster method at the singles and doubles (RCCSD) level to perform a systematic \emph{ab initio} investigation of the hyperfine structure of the low-lying states in $^{69,71}$Ga. We first calculate the energies and the hyperfine constants $A$ and $B$ of the low-lying states, and validate the reliability of our method through comparison with available experimental and theoretical results. On this basis, we further compute the first-order magnetic octupole hyperfine constant $C$ and systematically evaluate the second-order corrections arising from off-diagonal hyperfine interactions. Finally, starting from the original experimental hyperfine intervals of Daly and Holloway~\cite{Daly1954}, we apply our \emph{ab initio} matrix elements to re-extract the hyperfine constants and the nuclear magnetic octupole moments, providing an independent \emph{ab initio} assessment.

The paper is organized as follows. Section~\ref{method} describes the relativistic coupled-cluster method and the theoretical framework of hyperfine interactions. Section~\ref{results} presents the calculated results and discussion. Section~\ref{sec:summary} summarizes the conclusions.

\section{Theoretical methods}\label{method}

\begin{figure}[htbp]
  \centering
  \includegraphics[width=0.95\columnwidth]{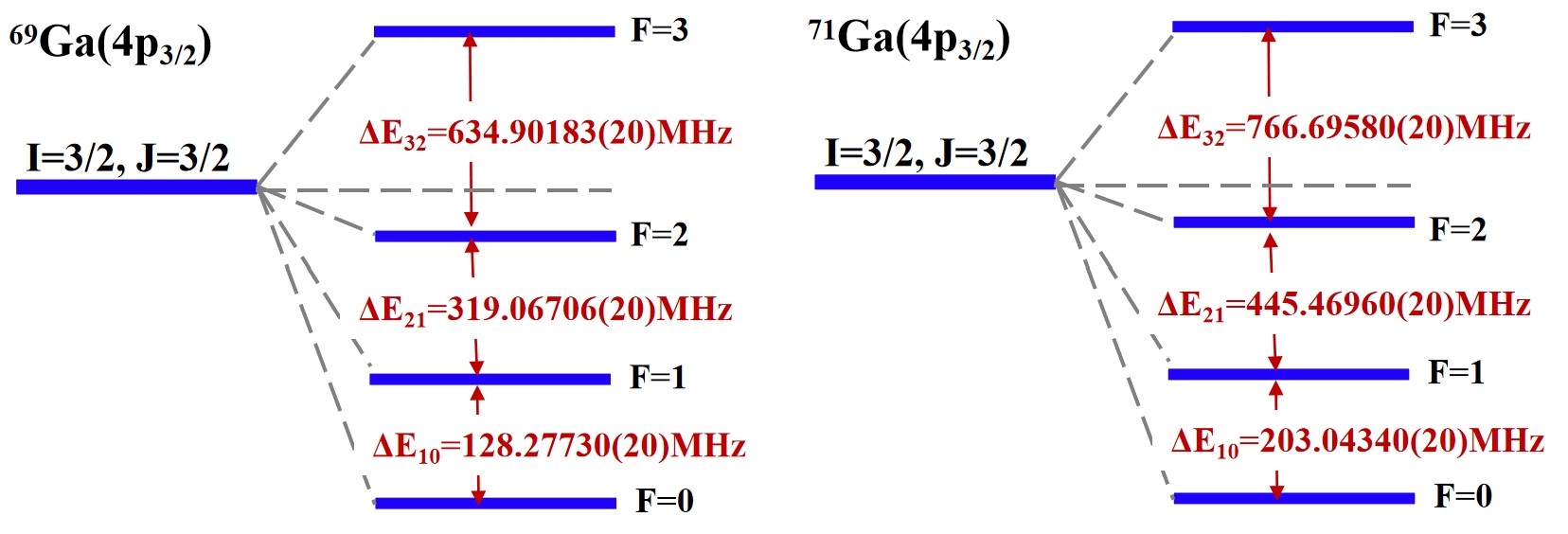}
  \caption{Hyperfine-structure splitting of the $4p_{3/2}$ state in $^{69}$Ga and $^{71}$Ga. The measured intervals (in MHz) are taken from Daly and Holloway~\cite{Daly1954}, with the uncertainties given in parentheses. The interval between the hyperfine levels $F$ and $F'$ is $\Delta E_{F'F}=E_{F'}-E_F$. The fine-structure level $J=3/2$ and the hyperfine levels $F=0$--$3$ are indicated.}
  \label{fig:hfs_levels}
\end{figure}

\subsection{Hyperfine-structure theory}

The hyperfine interaction couples the nuclear electromagnetic multipole moments with the electron cloud, splitting a fine-structure level $E_{J}$ into hyperfine levels $E_{F}$, where $\mathbf{F} = \mathbf{I} + \mathbf{J}$ is the total angular momentum of the atom, $\mathbf{I}$ the nuclear spin, and $\mathbf{J}$ the electronic angular momentum. Figure~\ref{fig:hfs_levels} illustrates the hyperfine splitting of the $4p_{3/2}$ state in $^{69}$Ga and $^{71}$Ga, together with the three measured intervals $\Delta E_{10}, \Delta E_{21}, \Delta E_{32}$ (in MHz, uncertainties in parentheses) taken from Daly and Holloway~\cite{Daly1954}.

When second-order hyperfine effects are included, the energy of a hyperfine level can be written as
\begin{equation}
E_{F} = E_{J} + E^{(1)}_{F} + E^{(2)}_{F},
\end{equation}
where $E^{(1)}_{F}$ is the first-order hyperfine correction, parameterized by the magnetic dipole ($M1$), electric quadrupole ($E2$), and magnetic octupole ($M3$) hyperfine constants $A$, $B$, and $C$:

\begin{small}
  \begin{equation}\begin{aligned}
  E^{(1)}_{F}=&\sum_{k_{1}=1}^{3}\langle (\gamma IJ) F M_{F}|T_{n}^{(k_{1})}\cdot T_{e}^{(k_{1})} |(\gamma IJ) F M_{F}\rangle\\
  =&\underbrace{\frac{1}{2}KA}_{M1:\,{k_{1}=1}}+\underbrace{\frac{1}{2}\frac{3K(K+1)-4I(I+1)J(J+1)}{2I(2I-1)2J(2J-1)}B}_{E2:\,{k_{1}=2}}\\
  &+\frac{1}{[I(I-1)(2 I-1) J(J-1)(2 J-1)]}\times\left\{(5/4){K}^{3}\right.\\
  &+5{K}^{2}+{K}\times[-3 I(I+1)
  \times J(J+1)+I(I+1)
  \\&\underbrace{+J(J+1)+3]-5 I(I+1) J(J+1)\}C\qquad\qquad}_{M3:\,{k_{1}=3}}.
  \end{aligned}
  \label{eq:firstorder}
  \end{equation}
  \end{small}

where $K = F(F+1)-I(I+1)-J(J+1)$. The constants $A$, $B$, $C$ are related to the nuclear moments and the reduced matrix elements of the hyperfine operators $T^{(k)}$ by
\begin{align}
A &=\frac{\mu}{I}\frac{\langle \gamma J\|T^{(1)}\|\gamma J\rangle}{\sqrt{J(J+1)(2J+1)}}, \label{eq:4}\\
B &=2Q\left[\frac{2J(2J-1)}{(2J+1)(2J+2)(2J+3)}\right]^{1/2}\langle \gamma J\|T^{(2)}\|\gamma J\rangle, \label{eq:5}\\
C &=\Omega\left[\frac{J(2J-1)(J-1)}{(J+1)(J+2)(2J+1)(2J+3)}\right]^{1/2}\langle \gamma J\|T^{(3)}\|\gamma J\rangle. \label{eq:6}
\end{align}
Here $\mu$, $Q$, and $\Omega$ are the nuclear magnetic dipole, electric quadrupole, and magnetic octupole moments, respectively; $\langle \gamma J\|T^{(k)}\|\gamma J\rangle$ is the reduced matrix element of the spherical tensor operator of rank $k$ in the atomic state $|\gamma J\rangle$, and $\gamma$ denotes additional quantum numbers. For the $4p_{3/2}$ state with $J=3/2$, the selection rules of the hyperfine interaction restrict the multipole expansion to $k \leq 3$.

The term $E^{(2)}_{F}$ accounts for the second-order hyperfine correction, arising mainly from the off-diagonal coupling to nearby fine-structure levels via the hyperfine operators. For a single intermediate state $|\gamma' J'\rangle$, this correction can be parameterized by three second-order constants $\eta$, $\zeta$, and $\xi$ corresponding to the M1--M1, M1--E2, and E2--E2 channels:

\begin{small}
\begin{equation}
\begin{aligned}
E_F^{(2)}=&\underbrace{\sum_{J^{\prime}}\left\{\begin{array}{lll}
    F & J & I \\
    k_{1} & I & J^{\prime}
    \end{array}\right\}^{2} \eta}_{M1-M1:\,k_{1}=1}
    +\underbrace{\sum_{J^{\prime}}\left\{\begin{array}{lll}
    F & J & I \\
    k_{1} & I & J^{\prime}
    \end{array}\right\}\left\{\begin{array}{lll}
    F & J & I \\
    k_{2} & I & J^{\prime}
    \end{array}\right\} \zeta}_{M1-E2:\,k_{1}=1,k_{2}=2}\\
    &+\underbrace{\sum_{J^{\prime}}\left\{\begin{array}{lll}
      F & J & I \\
      k_{1} & I & J^{\prime}
      \end{array}\right\}^{2} \xi}_{E2-E2:\,k_{1}=2},
\end{aligned}
\label{eq:secondorder}
\end{equation}
\end{small}

The total second-order correction is obtained by summing over the relevant intermediate states. The second-order constants are defined as
\begin{align}
\eta &= \frac{(I+1)(2I+1)}{I}\,\mu^{2}\,
\frac{|\langle \gamma J'\|T^{(1)}\|\gamma J\rangle|^{2}}{E_{\gamma J}-E_{\gamma J'}}, \label{eq:8}\\
\zeta &= \frac{(I+1)(2I+1)}{I}\sqrt{\frac{2I+3}{2I-1}}\,
\mu Q \nonumber\\
&\quad \times
\frac{\langle \gamma J'\|T^{(1)}\|\gamma J\rangle\langle \gamma J'\|T^{(2)}\|\gamma J\rangle}{E_{\gamma J}-E_{\gamma J'}}, \label{eq:9}\\
\xi &= \frac{(I+1)(2I+1)(2I+3)}{4I(2I-1)}\,Q^{2}\,
\frac{|\langle \gamma J'\|T^{(2)}\|\gamma J\rangle|^{2}}{E_{\gamma J}-E_{\gamma J'}}. \label{eq:10}
\end{align}

The experimentally measured hyperfine intervals $\Delta E_{F'F}$ contain both first- and second-order effects. For a system with nuclear spin $I=3/2$ and electronic angular momentum $J=3/2$, the intervals $\Delta E_{10}$, $\Delta E_{21}$, $\Delta E_{32}$ are related to the hyperfine constants by
\begin{align}
A^{4p_{3/2}} &= \frac{1}{20}\Delta E_{10}+\frac{4}{25}\Delta E_{21}+\frac{21}{100}\Delta E_{32} \nonumber\\
&\quad +\frac{1}{180}\eta-\frac{\sqrt{5}}{750}\zeta+\frac{3}{500}\xi, \label{eq1} \\[4pt]
B^{4p_{3/2}} &= -\frac{1}{4}\Delta E_{10}-\frac{2}{5}\Delta E_{21}+\frac{7}{20}\Delta E_{32} \nonumber\\
&\quad +\frac{1}{30}\eta+\frac{\sqrt{5}}{100}\zeta, \\[4pt]
C^{4p_{3/2}} &= \frac{1}{80}\Delta E_{10}-\frac{1}{100}\Delta E_{21}+\frac{1}{400}\Delta E_{32} \nonumber\\
&\quad +\frac{\sqrt{5}}{2000}\zeta-\frac{1}{1000}\xi. \label{eq3}
\end{align}
As can be seen from Eqs.~\eqref{eq1}--\eqref{eq3}, the second-order constants $\eta$, $\zeta$, and $\xi$ must be supplied by atomic-structure calculations of the off-diagonal matrix elements in order to extract the pure first-order hyperfine constants $A$, $B$, and $C$. Conversely, once $C$ is determined experimentally, the nuclear magnetic octupole moment $\Omega$ is obtained through the diagonal matrix element $C/\Omega$ in Eq.~\eqref{eq:6}.

The single-particle reduced matrix elements of the operators $T^{(k)}$ are expressed in terms of the large and small radial components $P(r)$ and $Q(r)$ of the Dirac wavefunctions:
\begin{align}
\langle \kappa_{v}\|T^{(1)}\|\kappa_{w}\rangle &= -\langle -\kappa_{v}\|C^{(1)}\|\kappa_{w}\rangle (\kappa_{v}+\kappa_{w}) \notag\\
&\quad \times \int_{0}^{\infty} dr \frac{P_{v}(r)Q_{w}(r) + P_{w}(r)Q_{v}(r)}{r^2} \, F^{(1)}(r), \label{eq:88}\\
\langle \kappa_{v}\|T^{(2)}\|\kappa_{w}\rangle &= -\langle \kappa_{v}\|C^{(2)}\|\kappa_{w}\rangle \notag\\
&\quad \times \int_{0}^{\infty} dr \frac{P_{v}(r)P_{w}(r) + Q_{v}(r)Q_{w}(r)}{r^3} \, F^{(2)}(r), \\
\langle \kappa_{v}\|T^{(3)}\|\kappa_{w}\rangle &= -\frac{1}{3}\langle -\kappa_{v}\|C^{(3)}\|\kappa_{w}\rangle (\kappa_{v}+\kappa_{w}) \notag\\
&\quad \times \int_{0}^{\infty} dr \frac{P_{v}(r)Q_{w}(r) + P_{w}(r)Q_{v}(r)}{r^4} \, F^{(3)}(r). \label{eq:99}
\end{align}
Here $\kappa=\ell(\ell+1)-j(j+1)-1/4$ is the relativistic angular-momentum quantum number. The nuclear magnetization distribution function $F^{(k)}(r)$ is adopted from the uniformly magnetized and charged sphere model:
\begin{equation}
F^{(k)}(r) = 
\begin{cases} 
\left(\dfrac{r}{R_N}\right)^{2k+1}, & r \le R_N,\\[8pt]
1, & r > R_N,
\end{cases}
\label{eq:Fnuc}
\end{equation}
where $R_N = \sqrt{5/3}\,\langle r^2 \rangle^{1/2}$ is the nuclear radius, and $\langle r^2 \rangle^{1/2}$ is the root-mean-square charge radius of the nucleus.

The reduced matrix elements of the spherical harmonics are
\begin{align}
\langle\kappa_{v}\|C^{(k)}\|\kappa_{w}\rangle &= (-1)^{j_{v}+1/2}\sqrt{(2j_{v}+1)(2j_{w}+1)}\notag\\
&\quad\times\left\{\begin{array}{lll}
j_{v} & k & j_{w} \\
1/2 & 0 & -1/2
\end{array}\right\}\pi(\ell_v,k,\ell_w),
\end{align}
with $\pi(\ell_v,k,\ell_w)=1$ when $\ell_v+k+\ell_w$ is even and zero otherwise.

\subsection{Computational details}

To evaluate the diagonal and off-diagonal reduced matrix elements with high precision, we employ the relativistic coupled-cluster method restricted to single and double excitations (RCCSD), based on the no-pair Dirac--Coulomb Hamiltonian. The nucleus is modeled as a uniformly magnetized and charged sphere with the radius defined in Eq.~\eqref{eq:Fnuc}. Correlation effects are examined at three levels of approximation: Dirac--Fock (DF), linearized coupled-cluster (LCCSD), and full CCSD. A detailed description of the method can be found in our previous works on Fr, La$^{2+}$, Ra$^{+}$, Th$^{3+}$, and Cs~\cite{Tang2017,Lou2019,Li2021b,Li2021c,Li2021a,Li2023}.

The single-particle Dirac orbitals are expanded in a finite basis set of even-tempered Gaussian-type functions,
\[
G_i(r) = \mathcal{N}_i\, r^{\ell+1} e^{-\alpha_i r^{2}},
\]
where $\mathcal{N}_i$ is the normalization constant and $\alpha_i = \alpha_0\beta^{\,i-1}$. The parameters $\alpha_0$ and $\beta$ are optimized separately for each orbital symmetry. The basis set parameters are listed in Table~\ref{basis}, where $N$ is the number of basis functions per symmetry, and $N_c$ and $N_v$ are the numbers of core and virtual orbitals, respectively. All core electrons and virtual orbitals with energies below $20\,000$\,a.u.\ are included in the correlation calculations.

\begin{table}[htbp]
\centering
\caption{Parameters of the Gaussian basis set. $N$ is the number of basis functions for each symmetry; $N_c$ and $N_v$ are the numbers of core and virtual orbitals, respectively.}
\label{basis}
\begin{ruledtabular}
\begin{tabular}{lcccccccc}
 & $s$ & $p$ & $d$ & $f$ & $g$ & $h$ & $i$ & $k$ \\
\hline
$\alpha_0\times10^{3}$ & 2.5 & 2.5 & 6.5 & 7.5 & 15 & 15 & 15 & 15 \\
$\beta$ & 1.91 & 1.90 & 1.95 & 2.0 & 2.0 & 2.0 & 2.0 & 2.0 \\
$N$      & 35 & 30 & 25 & 20 & 15 & 10 & 10 & 10 \\
$N_c$    & 4 & 2  & 1  & 0  & 0  & 0  & 0  & 0  \\
$N_v$    & 24 & 24 & 21 & 20 & 15 & 10 & 10 & 10 \\
\end{tabular}
\end{ruledtabular}
\end{table}

\section{RESULTS AND DISCUSSION}\label{results}

In this section we present our calculated energies and the magnetic dipole and electric quadrupole hyperfine constants $A$ and $B$ for the low-lying states of $^{69}$Ga. These results are systematically compared with available experimental data and with previous high-precision calculations to establish the reliability of the employed relativistic coupled-cluster method. Using the same computational framework, we then evaluate the diagonal $C/\Omega$ factor and the off-diagonal hyperfine matrix elements that are required for the second-order hyperfine corrections. By applying our \emph{ab initio} matrix elements to the original measured hyperfine intervals of Daly and Holloway~\cite{Daly1954}, we re-extract the nuclear magnetic octupole moments of $^{69}$Ga and $^{71}$Ga and compare them with the values reported in 1954. Throughout our analysis the following nuclear moments are adopted: for $^{69}$Ga, $\mu = 2.01659(5)\,\mu_N$ and $Q = 0.171(2)$\,b; for $^{71}$Ga, $\mu = 2.56227(2)\,\mu_N$ and $Q = 0.107(1)$\,b~\cite{Stone2005,WebElements}.

\subsection{Energies}

\begin{table*}[htbp]
\centering
\small
\setlength{\tabcolsep}{4pt}
\caption{Comparison of calculated and experimental energy levels (in cm$^{-1}$) for low-lying states of Ga I. ${ E_{\rm DF}}$ denotes the lowest-order Dirac-Fock energy. ${ E_{\rm LCCSD}}$ and ${ E_{\rm CCSD}}$ are the energies obtained using LCCSD and CCSD approximations, respectively. Values in parentheses are percentage deviations from NIST values.}
\label{tab:energy}
\begin{ruledtabular}
\begin{tabular}{lccccc}
 Level&${ E_{\rm DF}}$& ${ E_{\rm LCCSD}}$ & ${ E_{\rm CCSD}}$ &${ E_{\rm SD}}$~\cite{Safronova2006}&${ E_{\rm Expt.}}$\cite{NIST_ASD} \\
\hline
$4p_{1/2}$  & $-43006.38$ (11.12\%) & $-48627.15$ (0.50\%) & $\mathbf{-48297.32}$ (0.19\%) & $-48487$ (0.21\%) & $-48387.63$ \\
$4p_{3/2}$  & $-42282.55$ (11.10\%) & $-47829.61$ (0.56\%) & $\mathbf{-47489.28}$ (0.15\%) & $-47672$ (0.23\%) & $-47561.44$ \\
$5s_{1/2}$  & $-21932.70$ (7.06\%)  & $-23843.47$ (1.04\%) & $\mathbf{-23559.52}$ (0.17\%) & $-23784$ (0.78\%) & $-23599.10$ \\
$5p_{1/2}$  & $-14480.78$ (5.62\%)  & $-15412.09$ (0.45\%) & $\mathbf{-15311.57}$ (0.21\%) & $-15371$ (0.18\%) & $-15343.58$ \\
$5p_{3/2}$  & $-14385.52$ (5.56\%)  & $-15300.87$ (0.45\%) & $\mathbf{-15202.17}$ (0.20\%) & $-15272$ (0.26\%) & $-15232.56$ \\
\end{tabular}
\end{ruledtabular}
\end{table*}

We calculated the energies of the $4p_{1/2,3/2}$, $5s_{1/2}$, and $5p_{1/2,3/2}$ states in Ga atom using the Dirac-Fock (DF), LCCSD, and CCSD methods. The predicted energies, labeled $E_{\rm DF}$, $E_{\rm LCCSD}$, and $E_{\rm CCSD}$, are listed in Table~\ref{tab:energy}, together with the SD results of Safronova \emph{et al.}~\cite{Safronova2006}. The values in parentheses indicate the percentage deviations from the NIST experimental values~\cite{NIST_ASD}, denoted as $E_{\rm NIST}$.

From Table~\ref{tab:energy}, one can observe the following: (i) There are clear differences between the DF and CCSD energies, indicating significant electron correlation effects that are not captured by the DF approximation. The largest deviation occurs for the $4p_{1/2,3/2}$ states, reaching about 11\%. (ii) The CCSD results deviate from the experimental values by no more than 0.21\% for all states, showing improvement over the LCCSD results. This suggests that the inclusion of nonlinear terms in the cluster operator is necessary for achieving more accurate energy levels. (iii) The CCSD results agree well with the NIST values for all states. The deviation for the $4p_{3/2}$ state, which is the focus of the subsequent hyperfine structure analysis, is 0.15\%. For the $5s_{1/2}$ state, the CCSD deviation is 0.17\%, while the SD method gives 0.78\%. For the $4p_{1/2}$ and $5p_{3/2}$ states, the CCSD deviations are also smaller than those from SD, whereas for the $5p_{1/2}$ state the two methods yield comparable deviations. These comparisons show that the CCSD method employed in this work provides a good description of the low-lying energy levels of Ga, establishing a reliable foundation for the subsequent extraction of the magnetic octupole moment from the hyperfine splitting of the $4p_{3/2}$ state.

\subsection{Hyperfine structure constants $A$}

\begin{table*}[htbp]
\centering
\small
\setlength{\tabcolsep}{3pt}
\caption{The HFS constants $A$ (in MHz) of $^{69}$Ga at different correlation levels. The numbers in parentheses are the uncertainties of the recommended values. Some other \emph{ab initio} theoretical and experimental results are also listed for comparison.}
\label{tab:hfs_A}
\begin{ruledtabular}
\begin{tabular}{lccccc}
Method & $4p_{1/2}$ & $4p_{3/2}$ & $5s_{1/2}$ & $5p_{1/2}$ & $5p_{3/2}$ \\
\midrule
${A_{\rm DF}}$ & 1013 & 184 & 615 & 131.1 & 24 \\
${A_{\rm LCCSD}}$& 1335 & 205 & 1094 & 147.2 & 60 \\
${A_{\rm CCSD}}$& 1346(17) & 195(10)& 1067(28) & 152(5) & 50(10) \\
${A_{\rm SD}}$~\cite{Safronova2006} & 1334 & 207.6 & 1114 &150.8 &55.02 \\
${A_{\rm MBPT}}$~\cite{Jonsson1985}& 1351.7 & 188.6 & & & \\
${A_{\rm Expt.}}$~\cite{Neijzen1980} & 1339.1(1.2) & & 1068.8(1.0) & & \\
${A_{\rm Expt.}}$~\cite{Daly1954} & & 190.8 & & & \\
${A_{\rm Expt.}}$~\cite{Cheal2010}  & & 191.5(9) & 1069.5(1.5) & & \\
${A_{\rm Expt.}}$~\cite{deGroote2017b} & & 189.8(13) & 1067.7(18) & & \\
\end{tabular}
\end{ruledtabular}
\end{table*}
We calculated the magnetic dipole hyperfine structure constants $A$ for the low-lying states of $^{69}$Ga using the DF, LCCSD, and CCSD methods. The results are listed in Table~\ref{tab:hfs_A} together with other reported theoretical and experimental values. To estimate the uncertainty of our CCSD results arising from the truncation of the coupled-cluster expansion, we adopt a conservative approach similar to that used in Ref.~\cite{Li2024} for indium: the uncertainty is taken as the larger of the magnitude of the nonlinear correlation contribution $|\text{LCCSD} - \text{CCSD}|$ and $5\%$ of the total correlation contribution $|\text{CCSD} - \text{DF}|$, which is intended to conservatively cover possible contributions from higher-order correlation effects such as triple excitations.

As can be seen from Table~\ref{tab:hfs_A}, the DF method systematically underestimates the $A$ constants for all states; the deviation reaches $42\%$ for the $5s_{1/2}$ state, indicating that electron correlation effects play a crucial role in enhancing the electron density at the nucleus. The LCCSD results show a significant improvement over DF, and the inclusion of nonlinear terms in the cluster operator (CCSD) further refines the results, but the magnitude and sign of the nonlinear correction depend strongly on the specific state. For instance, for the $4p_{3/2}$ state, the LCCSD value is $205$\,MHz, while the CCSD result is $195(10)$\,MHz, giving a nonlinear correction of about $10$\,MHz, which shows that this state is rather sensitive to nonlinear correlations. In contrast, the nonlinear correction for the $4p_{1/2}$ state is only about $1\%$, almost negligible. Similarly, the $5p_{3/2}$ state exhibits a nonlinear correction of $10$\,MHz, highlighting the importance of nonlinear correlation effects for this state.

Comparison with experimental data further confirms the reliability of the CCSD method. The SD approach employed by Safronova~\emph{et al.}~\cite{Safronova2006} is equivalent to our LCCSD method in the coupled-cluster framework, and the two DF values are in close agreement. For the $4p_{3/2}$ state, our CCSD result of $195(10)$\,MHz agrees with the 2017 experimental value of $189.8(13)$\,MHz within the estimated uncertainty, and is closer to experiment than the SD/LCCSD results ($207.6$\,MHz and $205$\,MHz). The MBPT result of J\"onsson \emph{et al.}~($188.6$\,MHz) also shows good agreement with experiment. For the $5s_{1/2}$ state, the CCSD value of $1067(28)$\,MHz deviates from the experimental measurement $1068.8(1.0)$\,MHz by only about $0.16\%$, whereas the SD and LCCSD methods exhibit deviations of $4.3\%$ and $2.4\%$, respectively, clearly demonstrating the necessity of nonlinear correlation corrections for obtaining highly accurate $A$ constants. The small difference between the SD and LCCSD results may partly originate from the different basis sets employed---Safronova \emph{et al.} used a maximum partial wave of $l=6$, while the present work extends it to $l=7$, providing a more complete treatment of correlation.

For the $5p_{1/2}$ and $5p_{3/2}$ states, for which no experimental data are currently available, our CCSD predictions are $152(5)$\,MHz and $50(10)$\,MHz, respectively. The SD results of Safronova \emph{et al.}~($150.8$\,MHz and $55.02$\,MHz) fall within the uncertainty ranges of our CCSD values.

In summary, by comparing the DF, LCCSD, CCSD, and literature SD results, one can clearly see the stepwise contributions of electron correlation and the important role of nonlinear terms. The CCSD method employed in this work produces $A$ constants that agree with experiment within the estimated uncertainties for those states where measurements exist, providing a reliable basis for the subsequent extraction of the magnetic octupole moment.

\subsection{Hyperfine structure constants $B$}

\begin{table}[htbp]
\small
\centering
\caption{The HFS constants $B$ (in MHz) for the $4p_{3/2}$ and $5p_{3/2}$ states of $^{69}$Ga at different correlation levels. Uncertainties for our CCSD results are given in parentheses.}
\label{tab:B}
\begin{ruledtabular}
\begin{tabular}{lcccccc}
Level &${B_{\rm DF}}$ &${B_{\rm LCCSD}}$ & ${B_{\rm CCSD}}$ &${B_{\rm MCDF^a}}$ & ${B_{\rm Expt.}}$~\cite{Cheal2010} & ${B_{\rm Expt.}}$~\cite{deGroote2017b} \\
\midrule
$4p_{3/2}$ & 43.7 & 61.7 & 62.2(1.0) & 59.37 & 63(2) & 61(8) \\
$5p_{3/2}$ & 5.8 & 8.6 & 8.3(0.3) &  &  &  \\
\end{tabular}
\end{ruledtabular}
\par\smallskip
$^a$ Obtained from the MCDF calculation of Dong et al.\ (2012) for $^{71}$Ga~\cite{Wang2012a, Wang2012b}. The $B/Q$ factor was scaled by $Q(^{69}\mathrm{Ga})=0.171$\,b.
\end{table}

The electric quadrupole hyperfine constants $B$ for the $4p_{3/2}$ and $5p_{3/2}$ states of $^{69}$Ga are listed in Table~\ref{tab:B} together with other reported theoretical and experimental values. As can be seen from Table~\ref{tab:B}, electron correlation effects are substantial: from DF to CCSD, the $B$ value for $4p_{3/2}$ increases from 43.7 to 62.2\,MHz (a relative change of about 42\%), while for $5p_{3/2}$ it increases from 5.8 to 8.3\,MHz (about 44\%). The nonlinear contributions (LCCSD to CCSD) are 0.5\,MHz and 0.3\,MHz, respectively, indicating that their importance depends on the state. For the $4p_{3/2}$ state, our CCSD result of 62.2(1.0)\,MHz is in good agreement with the experimental values of 63(2)\,MHz~\cite{Cheal2010} and 61(8)\,MHz~\cite{deGroote2017b}. The MCDF value of 59.37\,MHz, obtained from the $^{71}$Ga calculation of Dong et al.\ by scaling the $B/Q$ factor, differs from our CCSD result by about 2.9\,MHz. For the $5p_{3/2}$ state, no experimental data are available; our CCSD prediction of 8.3(0.3)\,MHz may serve as a reference for future measurements.

\subsection{Magnetic octupole moment}

The magnetic octupole hyperfine constant $C$ is connected to the nuclear magnetic octupole moment $\Omega$ through the reduced matrix element given in Eq.~\eqref{eq:6}. To extract $\Omega$ reliably, both the diagonal $C/\Omega$ factor and the off-diagonal matrix elements that govern the second-order hyperfine corrections are required. Table~\ref{tab:Ga_matrix} collects these quantities for the $4p_{3/2}$ state, evaluated at the Dirac--Fock (DF), linearized coupled-cluster (LCCSD), and full coupled-cluster (CCSD) levels.

\begin{table}[htbp]
\centering
\caption{$C/\Omega$ in kHz/($\mu_N \times b$) and off-diagonal matrix elements in MHz from DF, LCCSD, and CCSD calculations. The uncertainties of the CCSD results are given in parentheses.}
\label{tab:Ga_matrix}
\begin{ruledtabular}
\begin{tabular}{lcccc}
Level & DF & LCCSD & CCSD & Final \\
\midrule
\multicolumn{5}{c}{$C/\Omega$ in kHz/($\mu_N\times b$)} \\
$4p_{3/2}$ & 0.499 & 0.714 & 0.711 & 0.711(11) \\
\midrule
\multicolumn{5}{c}{Off-diagonal matrix elements in MHz} \\
$\langle 4p_{3/2}||O^{(1)}||4p_{1/2}\rangle$ & $-151$ & $-243$ & $-273$ & $-273(30)$ \\
$\langle 4p_{3/2}||O^{(2)}||4p_{1/2}\rangle$ & $-594$ & $-834$ & $-842$ & $-842(13)$ \\
$\langle 4p_{3/2}||O^{(1)}||5p_{1/2}\rangle$ & $-54$ & $115$ & $54$ & $54(54)$ \\
$\langle 4p_{3/2}||O^{(2)}||5p_{1/2}\rangle$ & $-213$ & $-299$ & $-297$ & $-297(5)$ \\
$\langle 4p_{3/2}||O^{(1)}||5p_{3/2}\rangle$ & $-192$ & $-443$ & $-374$ & $-374(70)$ \\
$\langle 4p_{3/2}||O^{(2)}||5p_{3/2}\rangle$ & $-207$ & $-291$ & $-290$ & $-290(5)$ \\
\end{tabular}
\end{ruledtabular}
\end{table}

For the diagonal $C/\Omega$ factor, the DF value is $0.499$\,kHz/($\mu_N\times b$), while the CCSD result reaches $0.711$\,kHz/($\mu_N\times b$), an enhancement of about $43\%$, indicating a strong influence of electron correlation. The nonlinear terms (LCCSD $\to$ CCSD) contribute merely $0.3\%$, showing that the CCSD method is well converged. The recommended value is $0.711(11)$\,kHz/($\mu_N\times b$), where the uncertainty is taken as the larger of the LCCSD--CCSD difference and $5\%$ of the total correlation correction.

The off-diagonal matrix elements exhibit pronounced state- and operator-dependent correlation effects. The magnetic-dipole matrix element $\langle 4p_{3/2}\|O^{(1)}\|4p_{1/2}\rangle$ increases from $-151$\,MHz at the DF level to $-273$\,MHz at the CCSD level (an enhancement of about $80\%$), with the final value of $-273(30)$\,MHz. The electric-quadrupole matrix element $\langle 4p_{3/2}\|O^{(2)}\|4p_{1/2}\rangle$ reaches $-842(13)$\,MHz, enhanced by about $42\%$ due to correlation. These two matrix elements provide the dominant contributions to the second-order corrections, and their uncertainties propagate directly into the hyperfine constants.

The matrix elements involving the $5p$ intermediate states display strong cancellation between different correlation contributions. For instance, $\langle 4p_{3/2}\|O^{(1)}\|5p_{1/2}\rangle$ changes sign from $-54$\,MHz at DF to $+115$\,MHz at LCCSD and then drops back to $54$\,MHz at CCSD, resulting in nearly complete cancellation. Because of this instability, a conservative $100\%$ uncertainty is assigned to the $5p$-related matrix elements. However, since the energy denominators for the $5p$ states are much larger than that for the $4p_{1/2}$ state, their final contributions to the second-order corrections are negligible; therefore only the $4p_{1/2}$ intermediate state is retained in the following analysis.

Using the off-diagonal matrix elements and the nuclear magnetic-dipole and electric-quadrupole moments of $^{69}$Ga and $^{71}$Ga, the second-order hyperfine constants $\eta$, $\zeta$, and $\xi$ are calculated via Eqs.~\eqref{eq:8}--\eqref{eq:10}. For $^{69}$Ga we obtain $\eta = 0.082(18)$\,MHz, $\zeta = 0.0370(47)$\,MHz, and $\xi = 0.00419(55)$\,MHz; for $^{71}$Ga the corresponding values are $\eta = 0.132(29)$\,MHz, $\zeta = 0.0291(88)$\,MHz, and $\xi = 0.00161(91)$\,MHz.
With these second-order constants, the second-order corrections to the hyperfine intervals are evaluated using Eqs.~\eqref{eq1}--\eqref{eq3}, and the first-order hyperfine constants $A$, $B$, and $C$ are obtained. Table~\ref{tab:Ga_ABC_corrected} lists the uncorrected and corrected values for the $4p_{3/2}$ state of $^{69}$Ga and $^{71}$Ga, together with the results of Daly and Holloway~\cite{Daly1954} for comparison.

\begin{table*}[htbp]
\centering
%\footnotesize
\setlength{\tabcolsep}{3pt}
\caption{HFS constants $A$, $B$, and $C$ (in MHz) for the $4p_{3/2}$ state of $^{69}$Ga and $^{71}$Ga without and with second-order corrections from the $4p_{1/2}$ intermediate state. The column ``D\&H'' lists the values reported by Daly and Holloway~\cite{Daly1954} for comparison. Numbers in parentheses are $1\sigma$ uncertainties in units of the last quoted digit.}
\label{tab:Ga_ABC_corrected}
\begin{ruledtabular}
\begin{tabular}{crrrrrr}
\multicolumn{1}{c}{HFS} & \multicolumn{1}{c}{Uncorrected} & \multicolumn{1}{c}{Corr.\ M1--M1} & \multicolumn{1}{c}{Corr.\ M1--E2} & \multicolumn{1}{c}{Corr.\ E2--E2} & \multicolumn{1}{c}{Total} & \multicolumn{1}{c}{D\&H~\cite{Daly1954}} \\
\midrule
\multicolumn{7}{c}{$^{69}$Ga} \\[2pt]
$A$ & $190.79398(5)$  & $4.53(10)[-4]$   & $-1.10(14)[-4]$  & $2.51(33)[-5]$   & $190.79435(11)$ & $190.79428(15)$ \\
$B$ & $62.51949(12)$ & $2.72(60)[-3]$   & $8.26(106)[-4]$  & \multicolumn{1}{c}{0} & $62.52304(62)$ & $62.52247(30)$ \\
$C$ & $5.02(32)[-5]$ & \multicolumn{1}{c}{0} & $4.13(53)[-5]$   & $-4.18(56)[-6]$  & $8.73(62)[-5]$ & $8.4(6)[-5]$ \\[6pt]
\multicolumn{7}{c}{$^{71}$Ga} \\[2pt]
$A$ & $242.43342(5)$  & $7.3(16)[-4]$    & $-0.88(27)[-4]$  & $0.98(56)[-5]$   & $242.43408(18)$ & $242.43395(20)$ \\
$B$ & $39.39484(12)$ & $4.4(10)[-3]$    & $6.6(20)[-4]$    & \multicolumn{1}{c}{0} & $39.3999(10)$   & $39.39904(40)$ \\
$C$ & $8.60(32)[-5]$ & \multicolumn{1}{c}{0} & $3.3(10)[-5]$    & $-1.6(9)[-6]$    & $1.17(10)[-4]$  & $1.15(7)[-4]$ \\
\end{tabular}
\end{ruledtabular}
\par\smallskip
$[n]$ denotes multiplication by $10^{n}$, e.g., $4.53(10)[-4] = (4.53 \pm 0.10)\times10^{-4}$.\\
The D\&H $C$ values were converted from cps to MHz: $84(6)\,\mathrm{cps}=8.4(6)\times10^{-5}\,\mathrm{MHz}$ for $^{69}$Ga, and $115(7)\,\mathrm{cps}=1.15(7)\times10^{-4}\,\mathrm{MHz}$ for $^{71}$Ga.
\end{table*}
For $^{69}$Ga, the second-order corrections are dominated by the M1--M1 channel: $+4.53(10)\times10^{-4}$\,MHz for $A$ and $+2.72(60)\times10^{-3}$\,MHz for $B$, while the E2--E2 correction is negligible. The corrected $A$ and $B$ agree with the Daly and Holloway values within the error bars, with the uncertainties now dominated by the M1--M1 term rather than by the experimental interval error.

The situation is more pronounced for the $C$ constant. The uncorrected $C$ value is only $5.02(32)\times10^{-5}$\,MHz, much smaller than the $8.4(6)\times10^{-5}$\,MHz reported by Daly and Holloway. After adding the M1--E2 and E2--E2 corrections, the total $C$ becomes $8.73(62)\times10^{-5}$\,MHz, which agrees with the Daly and Holloway result to within about $3\times10^{-5}$\,MHz, well inside the combined uncertainties. The same holds for $^{71}$Ga: the corrected $C$ value is $1.17(10)\times10^{-4}$\,MHz, consistent with $1.15(7)\times10^{-4}$\,MHz.

This comparison demonstrates that the net second-order correction estimated by Daly and Holloway using the semiempirical Schwartz theory is almost identical to our \emph{ab initio} RCCSD result, confirming that their treatment of the $4p_{1/2}$ perturbation was physically sound. Building on this, the present study provides, for the first time, a clear separation and quantification of the individual M1--M1, M1--E2, and E2--E2 channels. The relative uncertainty of the total $C$ constant (about $7\%$) is appreciably larger than those of $A$ and $B$, because $C$ itself is tiny and the correction uncertainties propagate significantly into the final result.

Dividing the corrected $C$ values by $C/\Omega = 0.711(11)$\,kHz/($\mu_N\times b$) yields our recommended nuclear magnetic octupole moments:
\begin{align}
\Omega(^{69}\mathrm{Ga}) &= 0.123(9)\ \mu_N\times b, \label{eq:Omega69}\\
\Omega(^{71}\mathrm{Ga}) &= 0.165(15)\ \mu_N\times b. \label{eq:Omega71}
\end{align}

Compared with the original values $\Omega_{\mathrm{DH}}(^{69}\mathrm{Ga})=0.107(20)\,\mu_N\times b$ and $\Omega_{\mathrm{DH}}(^{71}\mathrm{Ga})=0.146(20)\,\mu_N\times b$ of Daly and Holloway, our central values are about $13\%$ and $11.5\%$ larger, respectively.

To identify the origin of the differences, the total relative deviation is decomposed into two contributions, as shown in Table~\ref{tab:Omega_compare_decomp}. If we keep the $C$ constants of Daly and Holloway and replace their implicit $C/\Omega$ factor (about $0.79$\,kHz/($\mu_N\times b$)) with our \emph{ab initio} value $0.711(11)$\,kHz/($\mu_N\times b$), $\Omega$ increases by $9.0\%$--$9.5\%$. This part originates entirely from the improved description of the electron density in the nuclear region. When we further replace the $C$ constants with our own corrected values, the octupole moments increase by an additional $2.0\%$--$4.0\%$. The two contributions add up to the total differences of $13.0\%$ and $11.5\%$.

\begin{table*}[htbp]
  \centering
  %\small
  \setlength{\tabcolsep}{4pt}
  \caption{Comparison of magnetic octupole moments and decomposition of differences between the present work and Daly and Holloway (1954)~\cite{Daly1954}.}
  \label{tab:Omega_compare_decomp}
  \begin{ruledtabular}
  \begin{tabular}{lcccccc}

  \multirow{2}{*}{Iso.} & \multicolumn{2}{c}{$\Omega$ ($\mu_N\times b$)} && \multicolumn{3}{c}{Difference relative to this work} \\
  \cmidrule{2-3}\cmidrule{5-7}
                           & This work & D\&H~\cite{Daly1954} & &Total & due to $C/\Omega$ & due to 2nd-order \\
  \midrule
  $^{69}$Ga & $0.123(9)$ & $0.107(20)$ & &$13.0\%$ & $9.0\%$ & $4.0\%$ \\
  $^{71}$Ga & $0.165(15)$ & $0.146(20)$ & &$11.5\%$ & $9.5\%$ & $2.0\%$ \\
  \end{tabular}
  \end{ruledtabular}
  \par\smallskip
  The ``due to $C/\Omega$'' column shows the relative change caused by replacing the $C/\Omega$ factor of Daly and Holloway (implicitly $\approx 0.79$~kHz/($\mu_N\times b$)) with our CCSD value $0.711(11)$~kHz/($\mu_N\times b$), while keeping their $C$ constant. The ``due to 2nd-order'' column reflects the effect of re‑evaluating the second‑order hyperfine corrections with our RCCSD method. The two contributions add up to the total difference.
\end{table*}
This decomposition clearly shows that the dominant source of the revision is the improved diagonal matrix element $C/\Omega$. The net second-order correction evaluated in this work is numerically close to that of Daly and Holloway, but a direct channel-by-channel comparison is not possible since the individual contributions were not reported in their work. Their octupole moments were systematically smaller because the semiempirical $C/\Omega$ factor available at that time overestimated the true value.

The current $\Omega$ uncertainties of $7\%$--$9\%$ are dominated by the second-order corrections. A further reduction would most effectively be achieved by improving the accuracy of the $\langle 4p_{3/2}\|O^{(1)}\|4p_{1/2}\rangle$ and $\langle 4p_{3/2}\|O^{(2)}\|4p_{1/2}\rangle$ matrix elements. The $C/\Omega$ factor and the second-order constants $\eta$, $\zeta$, and $\xi$ provided in this work can serve as a set of calibration parameters. If more precise measurements of the $4p_{3/2}$ hyperfine splitting become available in the future, they can be directly combined with the present framework to obtain improved nuclear magnetic octupole moments, providing a solid atomic-physics input for nuclear-structure studies of gallium isotopes.

\section{Summary}
\label{sec:summary}

In this work, we have performed a systematic \emph{ab initio} study of the hyperfine-structure properties of the low-lying states in $^{69}$Ga and $^{71}$Ga using the relativistic coupled-cluster method at the singles and doubles (CCSD) level. The reliability of the present approach is established through a detailed comparison of the calculated excitation energies, magnetic dipole hyperfine constants $A$, and electric quadrupole hyperfine constants $B$ with available experimental data and with previous high-precision calculations. For all considered states, our CCSD energies agree with the NIST recommended values to within $0.2\%$, and the $A$ and $B$ constants reproduce the most accurate measurements to within a few percent. This level of agreement confirms that the employed method captures the essential electron-correlation effects and provides a faithful description of the atomic wave functions in the nuclear region.

To extract the nuclear magnetic octupole moments, we have recalculated the diagonal $C/\Omega$ factor and the relevant off-diagonal hyperfine matrix elements that govern the second-order hyperfine corrections. By applying these matrix elements to the original measured hyperfine intervals of Daly and Holloway (1954), we obtain the following refined magnetic octupole moments:
$\Omega(^{69}\mathrm{Ga})$ = 0.123(9)\;$\mu_N\times b$, 
$\Omega(^{71}\mathrm{Ga})$ = 0.165(15)\;$\mu_N\times b$.
These values are approximately $13\%$ and $12\%$ larger, respectively, than those originally reported by Daly and Holloway. A quantitative decomposition of the differences shows that the dominant source of the revision is the improved diagonal matrix element $C/\Omega$, which benefits from the more realistic description of the electron density in the nuclear region provided by the present work. The net second-order correction evaluated in this work is numerically close to the semiempirical result of Daly and Holloway (1954), but a direct comparison at the channel level is not possible since the latter did not report the individual contributions. The present study, for the first time, separates and quantifies the contributions of the individual second-order channels based on \emph{ab initio} calculations, providing a more solid theoretical basis for the extraction of the magnetic octupole moment.

Finally, we emphasize that the present analysis can be straightforwardly updated when more precise hyperfine-splitting measurements become available; the $C/\Omega$ factor and the second-order constants provided in this work can then be directly combined with new experimental intervals to obtain improved nuclear magnetic octupole moments. The refined octupole moments, together with the accompanying \emph{ab initio} matrix elements, offer valuable benchmarks for nuclear-structure models and for future experimental investigations of higher-order electromagnetic moments in gallium and neighboring nuclei.

\begin{acknowledgments}
The work was supported by the National Natural Science Foundation of China under Grant No.12304269 and No.12174268, and the Launching Fund of Henan University of Technology (31401512).
\end{acknowledgments}

\bibliography{Ga}

\end{document}